# Critical review of patient outcome study in head and neck cancer radiotherapy

**Authors:**

Jingyuan Chen, PhD[1], Yunze Yang, PhD[2], Chenbin Liu, PhD[3], Hongying Feng, PhD[1,4,5], Jason M. Holmes, PhD[1], Lian Zhang, PhD[1,6], Steven J. Frank, MD[7], Charles B. Simone II, MD[8], Daniel J. Ma, MD[9], Samir H. Patel, MD[1], Wei Liu, PhD[1]

[1]Department of Radiation Oncology, Mayo Clinic, Phoenix, AZ 85054, USA

[2]Department of Radiation Oncology, the University of Miami, FL 33136, USA

[3]Cancer Hospital & Shenzhen Hospital, Chinese Academy of Medical Sciences and Peking Union Medical College, Shenzhen, China

[4]College of Mechanical and Power Engineering, China Three Gorges University, Yichang, Hubei 443002, People's Republic of China

[5]Department of Radiation Oncology, Guangzhou Concord Cancer Center, Guangzhou, Guangdong, 510555, People's Republic of China

[6]Department of Oncology, The First Hospital of Hebei Medical University, Shijiazhuang, Hebei, 050023, People's Republic of China

[7]Department of Radiation Oncology, The University of Texas MD Anderson Cancer Center, Houston, TX 77030, USA

[8]New York Proton Center, New York, NY 10035, USA

[9]Department of Radiation Oncology, Mayo Clinic, Rochester, MN 55905, USA

Corresponding author: Wei Liu, PhD, Professor of Radiation Oncology, Department of Radiation Oncology, Mayo Clinic Arizona; E-mail: Liu.Wei@mayo.edu.

**Conflicts of Interest Disclosure Statement**

No

**Acknowledgments**

This research was supported by the National Cancer Institute (NCI) R01CA280134, the Eric & Wendy Schmidt Fund for AI Research & Innovation, The Fred C. and Katherine B. Anderson Foundation, and the Kemper Marley Foundation.

**Data Availability Statement**

The data analyzed during the current study are not publicly available due to patient privacy concerns and institutional policies regarding protected health information (PHI). However, de-identified data that support the findings of this study are available from the corresponding author upon reasonable request and with appropriate institutional review board (IRB) approval.

**Ethical Approval**

This study was approved by Mayo Clinic Arizona institutional review board (IRB#: 24-011106).

**Abstract**

Rapid technological advances in radiation therapy have significantly improved dose delivery and tumor control for head and neck cancers. However, treatment-related toxicities caused by high-dose exposure to critical structures remain a significant clinical challenge, underscoring the need for accurate prediction of clinical outcomes—encompassing both tumor control and adverse events (AEs). This review critically evaluates the evolution of data-driven approaches in predicting patient outcomes in head and neck cancer patients treated with radiation therapy, from traditional dose-volume constraints to cutting-edge artificial intelligence (AI) and causal inference framework. The integration of linear energy transfer in patient outcomes study, which has uncovered critical mechanisms behind unexpected toxicity, was also introduced for proton therapy. Three transformative methodological advances are reviewed: radiomics, AI-based algorithms, and causal inference frameworks. While radiomics has enabled quantitative characterization of medical images, AI models have demonstrated superior capability than traditional models. However, the field faces significant challenges in translating statistical correlations from real-world data into interventional clinical insights. We highlight that how causal inference methods can bridge this gap by providing a rigorous framework for identifying treatment effects. Looking ahead, we envision that combining these complementary approaches, especially the interventional prediction models, will enable more personalized treatment strategies, ultimately improving both tumor control and quality of life for head and neck cancer patients treated with radiation therapy.

**Introduction**

More than 1 million new cancer cases are diagnosed, and more than 600,000 people die from cancer in the US every year[1, 2]. Radiotherapy (RT) plays a pivotal role in cancer treatment, being utilized in 50-75% of cancer patients either as primary treatment or in combination with other modalities[3]. Despite significant technological advancements in RT delivery systems, managing treatment-related adverse events (AEs) continues to pose substantial challenges across various disease sites[4]. Head and neck (H&N) cancer exemplifies the complexities inherent in modern radiotherapy[5-8]. Currently, most H&N cancer patients receive treatment through intensity-modulated radiotherapy (IMRT) or volumetric-modulated arc therapy (VMAT)[9]. The anatomical complexity of the H&N region presents unique challenges, as tumors often abut or infiltrate critical organs at risk (OARs), such as the brainstem, oral cavity, parotids, optic chiasm, etc.. This intricate anatomical relationship frequently results in high-dose radiation exposure to healthy tissues, leading to numerous potential AEs[10-16]. These complications can include brain necrosis[17, 18], dysphagia[19, 20], xerostomia[21, 22], laryngeal/pharyngeal edema[23] and so on. Such AEs often necessitate extensive supportive care both during and after treatment, significantly impacting patients' quality of life (QoL)[24]. Systematical research of AEs and accurate prediction of clinical outcomes in radiotherapy is essential for personalizing treatment strategies. Such predictions enable physicians to optimize the treatment plan - maximizing tumor control while minimizing damage to surrounding healthy tissues.

The digitization of healthcare data, particularly through electronic medical records (EMRs)[25], has enabled large-scale analysis of multi-modal datasets (imaging, radiomic, dosimetry and so on). These rich clinical repositories, combined with advances in artificial intelligence (AI), now empower researchers to identify subtle patterns across various clinical factors, medical imaging, dosimetric factors, and the patient outcomes[26-30]. Besides, the modern AI frameworks show the capacity to process complex, multi-dimensional data, that were previously out of the range of using conventional statistical methods[31, 32]. The synergy between healthcare digitization and AI has accelerated the development of robust outcome prediction models, enabling more personalized and evidence-based radiotherapy strategies.

In this comprehensive review, we examine the evolution and current state of data-driven patient outcome studies in radiation oncology. We begin by discussing conventional methods, including the Lyman-Kutcher-Burman (LKB) model and QUANTEC guidelines[33, 34], which have formed the foundation of patient outcome modeling. We then explore the emergence of radiomics-based approaches and the development of AI-based methods, which have revolutionized our ability to predict patient outcomes. Special attention is given to Linear Energy Transfer (LET)-related patient outcome studies in proton therapy, addressing the unique considerations in particle therapy. Finally, we discuss future directions in patient outcome research, particularly focusing on interventional individual patient outcome predictions, which promise to further advance the field of personalized radiation oncology.

**Lyman-Kutcher-Burman (LKB) model and QUANTEC guidelines**

The Lyman-Kutcher-Burman (LKB) model is a foundational framework in radiation oncology used to predict the normal tissue complication probability (NTCP) based on dose-volume histograms (DVHs)[35, 36]. LKB is a generalized linear model (GLM). It uses probit as the link function, which is the inverse of the cumulative distribution function (CDF) of the standard normal distribution. The general form can be written in terms of probability $p = \Pr(Y = 1|\mathbf{x})$

$$p = \Phi(\beta_0 + \beta_1 x_1 + \cdots + \beta_k x_k), \quad (1)$$

where $\beta_0, \beta_1, \ldots, \beta_k$ are the coefficients for the predictor variables $x_1, \ldots, x_k$.

$\Phi(\cdot)$ is the CDF of the standard normal distribution,

$$\Phi(z) = \frac{1}{\sqrt{2\pi}} \int_{-\infty}^{z} e^{-\frac{u^2}{2}} du, \quad (2)$$

In the context of LKB model of NTCP, the only predictor is often the effective dose $D_{\text{eff}}$, and the model can be written as

$$\text{NTCP} = \Phi\left(\frac{D_{\text{eff}} - TD_{50}}{m \times TD_{50}}\right) = \Phi\left(-\frac{1}{m} + \frac{1}{mTD_{50}} \times D_{\text{eff}}\right), \quad (3)$$

where $TD_{50}$ represents the tolerance dose of 50% complication probability, and $m$ describes the slope of the modeling curve at $TD_{50}$.

In the original work, Lyman used power law to determine the equivalent dose with partial volume irradiation treatment[36]. Building on this foundation, the generalized equivalent uniform dose (gEUD) concept was subsequently developed[37]. Under the power-law framework, it assumes that for each subunit of an organ irradiated with dose $D_i$ exhibits a local effect $E(D_i)$ follows the power law relationship:

$$E(D_i) = \left(\frac{D_i}{D_{ref}}\right)^{\frac{1}{n}}, \quad (4)$$

Where $D_{ref}$ is a presumable reference dose, $\frac{1}{n}$ is the power coefficient and $i = 1, 2, \cdots, k$ is the number of subunits.

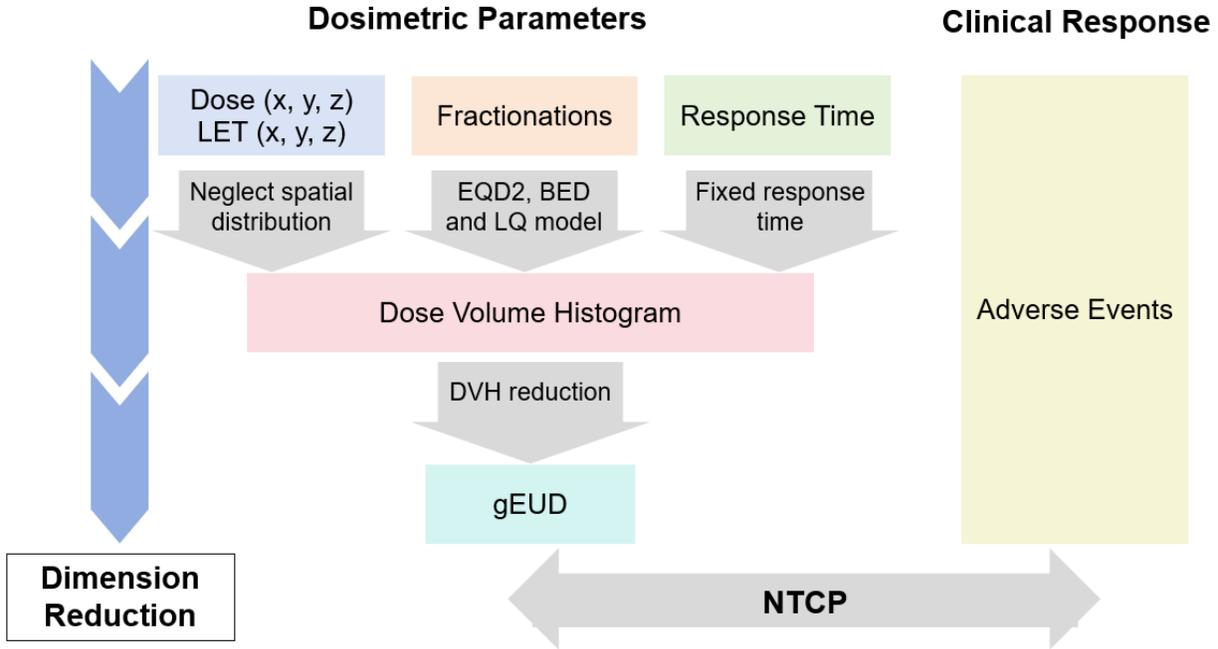

**Figure 1**. A conceptual framework in conventional normal tissue complication probability (NTCP) modeling. Dosimetric parameters are dimensionally reduced into one single variable, the generalized equivalent uniform dose (gEUD), for construction of the dose–response curve.

For tissues receiving heterogeneous irradiation, the relative damage volume (RDV) is defined as the sum of each subunit's dose-effect-weighted volume. Correspondingly, the uniform dose that produces the same RDV is referred to as the equivalent uniform dose (EUD)[38].

$$RDV = \sum_i v_i E(D_i), \quad (5)$$

$$EUD = E^{-1}\left(\sum_i v_i E(D_i)\right), \quad (6)$$

When the power-law assumption is applied, the generalized equivalent uniform dose (gEUD) can be derived[39].

$$RDV = \sum_i v_i \left(\frac{D_i}{D_{ref}}\right)^{\frac{1}{n}}, \quad (7)$$

$$gEUD = E^{-1}\left(\sum_i v_i \left(\frac{D_i}{D_{ref}}\right)^{\frac{1}{n}}\right) = \left(\sum_i v_i D_i^{\frac{1}{n}}\right)^n = \left(\sum_i v_i D_i^a\right)^{\frac{1}{a}}, \quad (8)$$

When each subunit is considered at the resolution of the dose grid, $E(D_i)$ acts as a weighting function of the dose bin $D_i$, which applies to its corresponding volume $v_i$[40]. The power coefficient, often denoted by $n$ or $a = 1/n$, adjusts how different dose levels are weighted in the overall calculation. For instance, when $a = 1$, the local effect is directly proportional to the received dose, causing the gEUD to reduce to the mean dose. In contrast, as $a$ approaches infinity, higher doses dominate, and the gEUD becomes increasingly influenced by the maximal dose. Conversely, for target volume, $a$ is often assumed to approach negative infinity, causing the gEUD become driven by the minimum dose. gEUD is commonly used in LKB model as the single predictor, referred to as the effective dose $D_{\text{eff}}$ in equation (3).

Complementing these predictive modeling efforts, the Quantitative Analysis of Normal Tissue Effects in the Clinic (QUANTEC) guidelines consolidate clinical data to provide dose tolerance thresholds for various organs and tissues[34]. Published in 2010, these guidelines are rooted in extensive review of existing literature and serve as an evidence-based reference to inform clinical decision-making[34]. QUANTEC guidelines include organ-specific recommendations that consider both partial- and whole-organ radiation exposures, enabling clinicians to design treatment plans that respect established safety thresholds.

Together, LKB model and QUANTEC guidelines underscore the importance of a quantitative, evidence-based approach to modern radiotherapy. A primary limitation of QUANTEC is that its recommendations largely stem from clinical data collected in 3D conformal radiation therapy settings, potentially restricting their applicability to more advanced modalities such as IMRT, VMAT, or Stereotactic Body Radiation Therapy (SBRT). While the LKB model offers a conceptual and straightforward mathematical framework for predicting complication rates, its principal drawback lies in the dimension reduction process: by collapsing the entire dose–volume distribution into a single metric, which can potentially overlook the spatial complexity and heterogeneity of dose delivery.

**Radiomics**

Radiomics[41], the high-throughput extraction of quantitative features from medical images, has emerged as a transformative tool in radiation oncology, particularly for predicting treatment response in patients undergoing radiotherapy. This advanced image analysis technique allows for a more comprehensive characterization of tumor heterogeneity, going beyond traditional imaging metrics such as size and volume[42-44]. This approach has shown significant potential in predicting tumor and normal tissue response to radiation treatment[45-47]. By quantifying intratumoral heterogeneity, texture analysis can provide valuable insights into tumor biology and treatment outcomes, potentially improving the accuracy of clinical outcome predictions[48, 49]. Figure 2 illustrates the comprehensive workflow of radiomics-driven precision in radiotherapy.

Delta radiomics, which involves analyzing changes in radiomic features over time, has gained attention for its ability to capture dynamic tumor characteristics during treatment[43]. This approach can potentially offer early indicators of treatment response, allowing for timely adjustments to therapy plans, but requires validation in multi-institutional trials addressing scanner variability. Additionally, multi-modal image radiomics, which integrates features from various imaging modalities (e.g., CT, MRI, PET, etc.), has shown promise in providing a more comprehensive tumor characterization and improving prediction accuracy[49,50].

Despite its potential, the field of radiomics faces several challenges that need to be addressed to enhance its clinical impact. These include the need for standardized image acquisition and reconstruction protocols, more accurate methods for region of interest identification, and robust feature selection techniques[50,51]. Overcoming these limitations will be crucial for establishing radiomics as a reliable tool for treatment response prediction in radiation oncology. As the field continues to evolve, the integration of radiomics with other advanced technologies, such as artificial intelligence and machine learning, holds promise for further improving the precision and personalization of radiotherapy[51,52].

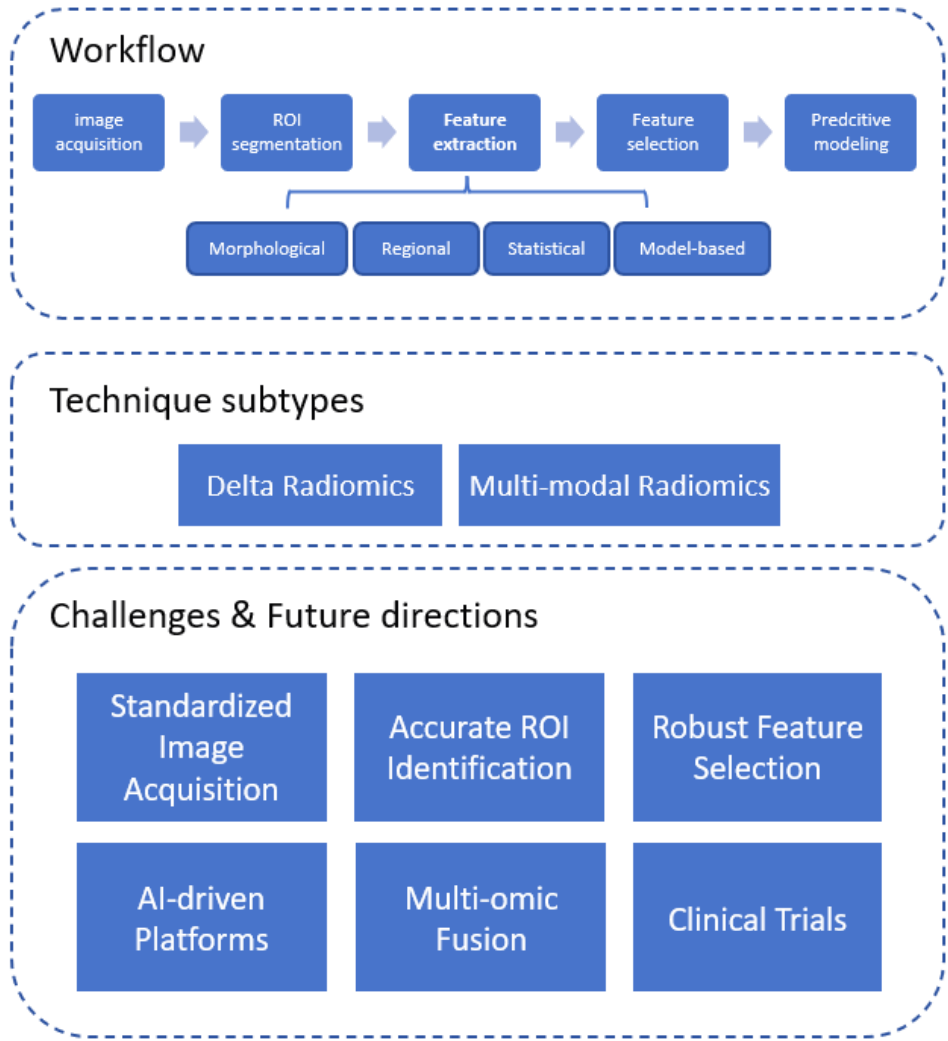

**Figure 2**. Radiomics-Driven Precision in Radiotherapy

**AI-based approaches in patient outcome studies**

In recent years, AI-based methods have been widely adopted for patient outcome prediction[53, 54], owing to their superior ability to represent data patterns compared to traditional approaches.

Machine learning (ML)[55], one of the core technologies within AI, has demonstrated remarkable capabilities in the medical domain[28, 56]. ML can construct predictive models by learning logical patterns from historical data to forecast patient outcomes[55]. Common ML approaches include linear models[57-61], decision tree models[62], ensemble models[55, 63-65] and so on . Linear models, such as logistic regression[57-59], linear discriminant analysis (LDA)[60], and support vector machines (SVM) with linear kernels[61], establish relationships between independent variables and outcomes through linear equations. Decision tree models like classification and regression trees (CART) capture nonlinear relationships through their branching structure[62]. Ensemble methods build upon the decision tree framework by generating predictions based on multiple tree models - for example, random forests[63, 64] combine multiple decision trees through bagging (bootstrap aggregating), while gradient boosted machines like XGBoost, light GBM, and CatBoost sequentially build trees to correct errors made by previous models[65]. ML methods can be categorized into supervised and unsupervised approaches. Supervised learning methods discussed above are particularly well-suited for developing patient outcome prediction models based on labeled data. In contrast, unsupervised learning methods, such as clustering, are valuable for revealing hidden data patterns and identifying potential clinical factors that may significantly impact patient outcomes[66].

ML has been extensively applied to improve prognosis in medical field[57, 60-62], particularly in oncology, where accurate prognostication is crucial for clinical purposes. ML methods have demonstrated improved accuracy in predicting cancer susceptibility, recurrence, and survival rates[67-69]. Researchers in biomedical and bioinformatics fields have leveraged ML tools to classify cancer patients into high- or low-risk recurrence groups for enhanced prognosis management and enabling better clinical management of patients[65, 70]. ML has become a cornerstone in patient outcome study in radiotherapy. ML models have been employed to analyze hypothyroidism complications caused by radiotherapy in H&N cancer patients[71], predict for pneumonitis and esophagitis followed radiotherapy for lung cancer[72, 73], analyze variations in relative biological effectiveness (RBE) values in proton therapy[61, 74-83], and develop extreme gradient boosting survival embeddings (XGBSE) for survival analysis[84, 85].

However, traditional ML methods rely on structured data, making it challenging to work with high-dimensional data such as medical images, dose distributions, and unstructured text records. Deep learning (DL), which utilizes multi-layered neural networks, has emerged as a powerful solution to overcome these limitations[86]. DL algorithms are particularly adept at processing large, high-dimensional datasets and have demonstrated remarkable success in tasks involving image analysis and unstructured text processing[87-99].

Several DL architectures have demonstrated significant potential in medical applications[100, 101]. Convolutional neural networks (CNNs) and U-Net have been extensively studied and widely applied to tasks such as CT image segmentation, dose distribution prediction, dose denoising, and synthetic image generation, etc. [93, 98, 99, 102-108]. Recurrent neural networks (RNNs) and long short-term memory (LSTM) networks have proven effective in analyzing sequential data including patient treatment histories and temporal outcomes[109, 110]. Additionally, generative adversarial networks (GANs) have been utilized for synthetic medical image generation and data augmentation[111-113].

In the field of patient outcome study in radiotherapy, DL approaches have shown promising results. CNNs have successfully predicted treatment outcomes in patients with H&N squamous cell carcinoma[114, 115]. CNN-based ensemble learning has enabled accurate prediction of brain metastasis (BM) local control outcomes after stereotactic radiosurgery (SRS) by analyzing CT images and dose maps[116]. Furthermore, 3D-ResNet architectures have demonstrated effectiveness in analyzing PET/CT imaging to predict distant metastasis (DM) and overall survival (OS) in H&N[117]. It is worth noting that DL typically requires significantly larger training datasets and greater computational resources compared to traditional ML methods to achieve optimal performance and generalization.

While DL approaches offer sophisticated modeling capabilities, they present challenges in interpretability. The CNN models are often referred to as "black box" methods due to the inherent difficulty in extracting meaningful insights from their decision-making processes[86, 118]. This opacity can hinder their adoption in clinical contexts, where understanding the rationale behind predictions is essential for trust and actionable insights.

In recent years, explanatory ML methods have significantly improved model interpretability[119, 120]. Among these, the Shapley additive explanations (SHAP) method has gained prominence as a tool to explain feature importance in prediction models[121, 122]. Rooted in game theory, SHAP methods assign quantified importance scores to individual features, offering granular insights into model behavior. As a model-agnostic framework, SHAP is compatible with many ML architectures, regardless of their internal structure and complexity. SHAP has been successfully applied in various radiation oncology studies[120, 123], such as identifying dosimetric predictors for toxicity in non-small cell lung cancer (NSCLC) patients[124] and evaluating factors affecting survival in nasopharyngeal cancer patients following radiotherapy[125].

Beyond post hoc tools like SHAP, inherently interpretable "white-box" models, such as Generalized Additive Models (GAM), and interpretable trees[126-128], provide direct interpretability. State-of-the-art examples include the explainable boosting machine (EBM)[129], which enhances GAM with cyclic gradient boosting and tree-based weak learners, and neural additive models (NAM), which train separate neural networks to model interpretable feature-specific relationships[130]. In radiotherapy, EBM has been used to identify dose-volume constraints (DVCs) on cardiopulmonary substructures (CPSs) associated with overall survival (OS) in NSCLC patients after radiotherapy[123, 124]. While white-box models may slightly underperform compared to cutting-edge black-box models in terms of model accuracy and capacity, their transparency and interpretability is indispensable for clinical applications requiring rigorous validation and intuitive explanations.

**Linear Energy Transfer (LET)-related patient outcome study in proton therapy**

Over the past few decades, proton therapy has experienced substantial technological advancements and expanded clinical applications[131-135]. A distinctive feature of the proton beam is its Bragg peak, characterized by a sharp dose drop-off beyond the target area. This characteristic creates a sharp dose reduction beyond the target region, potentially offering superior dose conformity and enhanced protection of surrounding healthy tissues compared to photon-based radiation therapy. These advantages have driven both technological advancement and broader clinical adoption over recent decades[82, 136-142].

However, proton therapy faces a significant challenge concerning RBE [74-79, 81, 143-146]. The interaction between protons and biological tissue is fundamentally different from photon radiation, as protons concentrate their energy deposition within a confined space. This concentrated energy results in elevated LET near the Bragg peak's distal end[143, 144, 147-150]. While laboratory studies consistently demonstrate increased RBE with higher LET values[151, 152], the clinical evidence remains less definitive, creating a gap between theoretical understanding and practical application[153-159].

The scientific community has documented various AEs potentially linked to RBE values exceeding the standard clinical assumption of 1.1. These include rib fractures[160, 161], rectal bleeding[162, 163], mandibular osteoradionecrosis[163, 164, 165], brain necrosis[153-159], and late-phase pulmonary changes[166] in proton therapy patients. Understanding the intricate relationships between physical dose, LET, and clinical outcomes has become crucial for optimizing proton therapy treatment planning. Current research approaches to studying LET-related clinical outcomes follow two main methodologies. The first employs voxel-based analysis, comparing affected and healthy tissue regions at a microscopic level. Such voxel-based analysis uses individual voxels as independent data points for regression analysis. This method, while detailed, relies on the presumed independence of voxels within affected regions and the attribution of damage solely to dosimetric factors. Clinical observations suggest that the assumption of voxel independence may not hold. Furthermore, radiation-affected regions can evolve over time as a result of ongoing biological processes. Here, the DLVH is emphasized as a novel tool for analyzing the joint distribution of Dose and LET. As shown in Figure 3, this tool serves as a three-dimensional upgrade to DVH and LETVH, providing an intuitive description of the joint Dose-LET distribution[162]. This tool has been applied in patient outcome study in H&N cancer and treatment plan optimization[162, 163, 167].

The second methodology examines outcomes at the organ level, comparing complications between photon and proton therapy patient populations. While this approach has identified potential increases in RBE related to proton therapy, quantifying LET's specific contribution remains challenging due to its heterogeneous distribution within tissues and complex relationship of tissue damage with dose distribution.

A promising new direction has emerged through dosimetric seed spot analysis, which introduces spatial clustering to identify independent lesion origins. This technique reduces biological noise and improves data independence by focusing on representative clusters rather than individual voxels. Recent studies utilizing this approach, particularly in H&N cases, suggest that current clinical practices may underestimate RBE values and that LET-enhancement plays a crucial role in initiating AEs[163, 164].

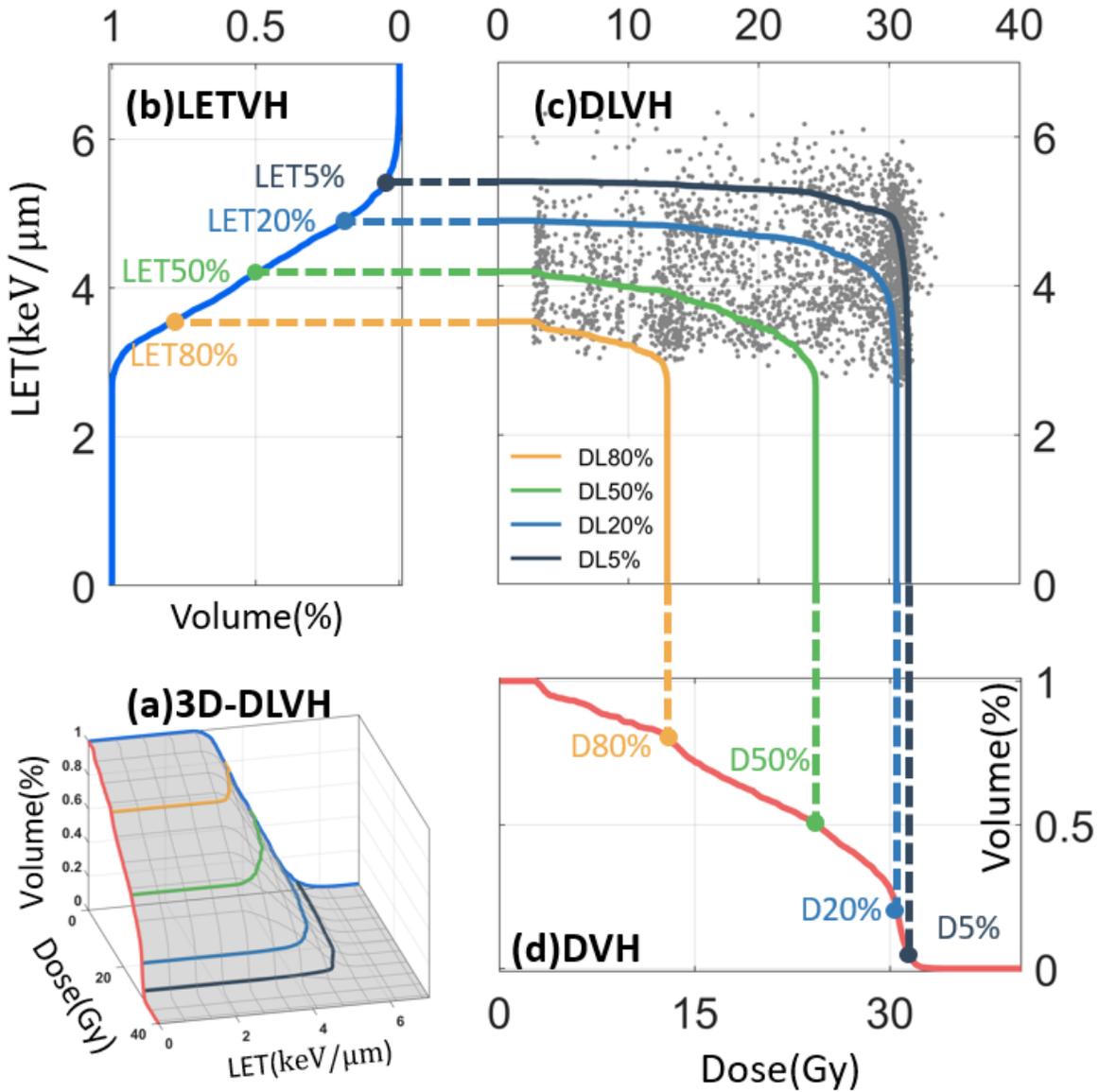

**Figure 3**. Sketches about the dose linear-energy-transfer (LET) volume histogram (DLVH). (a) Three-dimensional DLVH surface. The solid lines on the surface are the iso-volume contour lines DL$v$%. When viewing the 3D DLVH surface plot intersected with the Dose-Volume plane, we obtain the Dose-Volume Histogram (DVH), illustrated in panel (d).In panel (b), the intersection of the 3D-DLVH with the LET-Volume plane produces the LET-Volume Histogram (LETVH). Panel (c) shows the intersection of the 3D-DLVH with the Dose-Volume plane produces the Dose-

Volume Histogram (DVH).

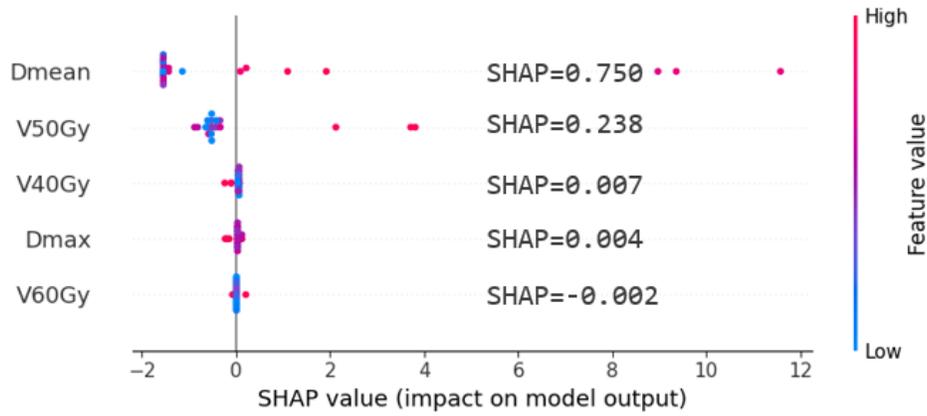

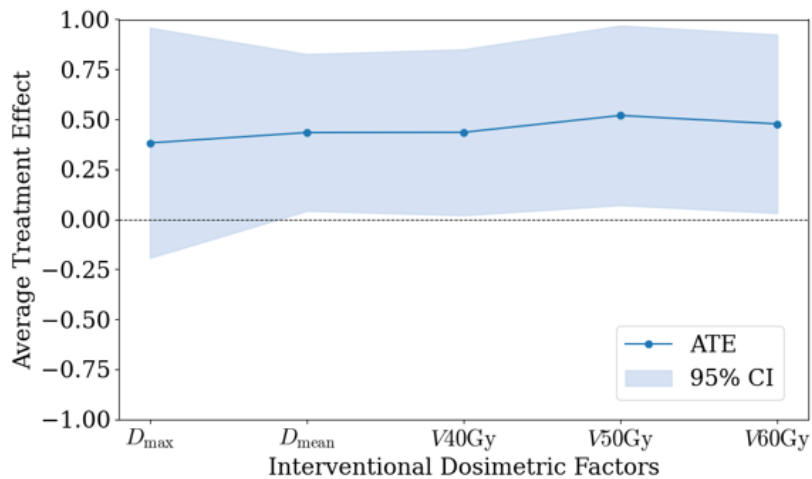

**Figure 4**. Comparison of interpretable machine learning and causal inference approaches in analyzing osteoradionecrosis (ORN) for head and neck radiation therapy in the same coherent. (a) SHAP values showing the impact of dosimetric factors on model predictions. Normalized SHAP values are noted. Only $D_{mean}$ and V50Gy are identified as important dosimetric factors (b) Causal inference results showing Average Treatment Effects (ATE) with 95% confidence intervals for different dosimetric factors. All important factors are identified as in previous study[167]. However, the causal inference approach requires more data to achieve a more accurate result.

**Interventional causal outcome prediction methods**

As outlined in previous sections, both traditional and interpretable ML models have advanced significantly. However, the correlation results derived from observational real-world data (RWD) require cautious interpretation, as they do not inherently reflect causal relationships[168]. For instance, SHAP-derived feature importance scores cannot be equated to causal effects—modifying a feature value does not guarantee a proportional change to the real outcome[127, 169]. Cooper et al. demonstrated these limitations of CNN and ML models based on real-world observational data through their related work[170, 171]. Their analysis using CNN and ML models on real-world data yielded a paradox: patients with asthma appeared to have a lower mortality risk when admitted with pneumonia. However, the lower observed mortality rate was actually due to hospital protocols - asthmatic patients with pneumonia typically receive more aggressive treatment and immediate intensive care, resulting in better survival rates compared to the general pneumonia patient population. This example highlights a critical limitation: making interventional recommendations based purely on statistical patterns in real-world data can lead to dangerously misleadingconclusions[169].

Given these challenges, establishing causal relationships--not just correlations--is essential when using patient outcome research to inform clinical practice[168, 169, 172]. Randomized controlled trials (RCTs) serve as the gold standard for establishing causal relationships, as random treatment allocation minimizes confounding bias[168, 169]. However, due to ethics constraints and the practical barrier associated with designing RCTs, the data from such trials is far more limited compared to RWD. Causal inference addresses this gap by providing a robust methodological framework to uncover causal relationships between treatments and outcomes, enabling reliable estimation of treatment effects using RWD[172-174].

Causal inference operates by estimating counterfactual outcomes - the unobservable outcomes that patients would experience under alternative treatments[169, 173]. By eliminating various biases present in RWD, this approach can identify true causal relationships between interventions and outcomes, forming the basis for actionable clinical recommendations[175]. Compared to traditional models, causal inference methods enable the estimation of individualized treatment effects and personalized predictions of potential patient outcomes under different treatment scenarios. The capability of causal inference models to estimate individual patient outcomes under various treatment options significantly advances the development of precision medicine models[176-178].

Recent years have seen notable developments in causal inference techniques[173]. These advances include generalized random forest-based and neural network-based estimators for multiple treatment effects[179], variable coefficient networks and generative adversarial models for continuous dosage-outcome estimation[180, 181], neural network-based multiple treatment estimators[182, 183], and federated learning approaches for individual dose-response functions that account for the joint effects of multiple dependent dosages[184]. Causal models are increasingly applied in healthcare. For instance, causal machine learning models quantify how interventions such as antidiabetic drugs alter diabetes risk, directly informing personalized treatment strategies[185, 186]. Figure 4 demonstrates the distinct performance of interpretable machine learning and causal inference approaches when applied to real-world biased data in radiation therapy outcomes. In our cohort study of 100 head and neck cancer patients who underwent photon therapy, 10 patients developed osteoradionecrosis (ORN) following treatment. We investigated the relationships between dosimetric parameters (Dmax, Dmean, V40Gy, V50Gy, and V60Gy) and ORN, under clinical factors including smoking history, hypertension, diabetes, chemotherapy, tumor type, dental extraction, and age. Due to the inherent selection bias in the distribution of clinical factors, traditional interpretable machine learning methods failed to accurately identify the crucial impact of V40Gy, V60Gy and $D_{max}$ dosimetric parameters on ORN development. In contrast, causal machine learning frameworks, specifically designed to handle biased observational data, successfully identified these critical dosimetric variables as significant predictors of ORN risk, highlighting the importance of causal approaches in radiation oncology research where treatment selection bias is prevalent.

However, applying causal models to radiotherapy patient outcome research still faces several challenges. First, causal inference relies on strict assumptions, including the confoundedness assumption, also known as strong ignobility, which assumes that all confounding variables that influence both interventions and outcomes are observed and measured. This is particularly difficult to verify in complex radiotherapy settings, where unmeasured confounders (e.g., subtle anatomical variations, treatment adherence) may persist. Second, estimating treatment effects through causal inference, particularly for individualized treatment effects, requires large data sizes to achieve reliable results with small uncertainties. Third, while high dimensional multi-modality data (e.g., CT scans and dose distributions) are critical for radiotherapy treatment planning and patient outcome prediction, causal algorithms capable of integrating and analyzing such high-dimensional, multimodal data remain underdeveloped. Current causal framework that can effectively incorporate and analyze such high dimensional data or synthesize multimodal data with clinical variables for robust causal reasoning are still in early stages of development.

**Conclusion**

The field of data-driven patient outcome studies has undergone transformative growth, driven by advancements in radiomics, AI/ML, and causal inference algorithms. AI/ML-powered approaches have demonstrated impressive accuracy in prediction tasks. These innovations have expanded the ability to analyze complex clinical data and derive meaningful insights for patient care. However, translating findings from real-world observational studies into clinical practice demands caution. Rigorous interpretation of results is critical, as statistical associations derived from observational data may reflect confounding rather than causation. Misattributing correlation to causality risks misguides clinical interventions, underscoring the need for robust causal frameworks to validate relationships between variables and outcomes.

Interventional patient outcome studies represent a promising direction for future research. By leveraging AI capabilities while properly accounting for causality, these approaches could revolutionize treatment planning and decision-making in personalized radiotherapy. Realizing this potential will require continued refinement of these methodologies, coupled with rigorous validation across diverse clinical cohorts and seamless integration into clinical workflows. Ultimately, harmonizing AI's analytical prowess with causal rigor will be key to transforming the deluge of clinical data into strategies that reliably improve patient outcomes.